# Discord-Enabled Teleportation-Inspired Optical Imaging at a Distance

Zilin Chen[1,†], Weiqian Shu[1,†], Xiaodong Qiu[2,*], Yangjian Cai[3,*], and Lixiang Chen[1,*]

[1]*Department of Physics, Xiamen University, Xiamen 361005, China*

[2]*Department of Electrical Engineering, City University of Hong Kong, Kowloon, Hong Kong SAR 999077, China*

[3]*School of Physics and Electronics, Shandong Normal University, Jinan 250014, China*

[*]*xiaodqiu@cityu.edu.hk*

[*]*yangjiancai@sdnu.edu.cn*

[*]*chenlx@xmu.edu.cn*

[†]*These authors contributed equally to this work*

**In quantum teleportation, a pair of entangled photons are prerequisite to serve as the quantum channel for quantum state transfer distantly. Here, we report a new strategy of quantum-teleportation-inspired classical optical imaging, which also works non-locally at a distance; however, only a classically correlated light source is used instead of entanglement. In our experiment, we explore the pseudo-thermal light source to offer the teleportation-like channel and employ the sum-frequency generation to perform the Bell-like state measurement. We successfully demonstrate the teleportation-inspired optical imaging of simple characters, Taiji diagram, and the superposition of orbital angular momentum modes. Moreover,**

**we experimentally observe that a better coherence of pseudo-thermal light will result in a lower contrast of the formed images, and thus revealing that non-zero quantum discord offered by pseudo-thermal light, regardless of zero entanglement, plays the pivotal role in sustaining the teleportation-like channel for imaging at a distance.**

One of the best signatures of nonclassicality in a quantum system is the existence of correlations that have no classical counterpart[1]. Entanglement is the most prominent of these correlations, which lies at the heart of the Einstein, Podolsky, and Rosen (EPR) paradox[2,3]. In entanglement, quantum states become correlated in such a way that the state of one particle is directly related to the state of the other, irrespective of the distance between them, which Einstein referred to as "spooky action at a distance." Quantum entanglement not only played a central role in the fundamental test of quantum mechanics, but also become a key resource that sparks a variety of quantum technological revolutions[4-7].

When quantum correlation is mentioned, we might be immediately tempted to think of entanglement. But this thinking has been gradually changed since the discovery of quantum discord[8-9], which describes the fact that some quantum systems even without entanglement are capable of possessing nonclassical correlations. Discord was proposed as a figure of merit for characterizing the nonclassical resource in the quantum vs classical scenario[8-9]. It was even believed that entanglement is the key resource that possesses the quantum-enhanced advantage[10]. However, this brief also started to change, as discord has been explored to provide enhancement for a variety of quantum tasks, such as deterministic quantum computation[11,12], quantum state preparation[13], quantum state emerging[14,15], quantum state broadcasting[16], and discord-consumed information encoding[17]. Therefore, it is suggested that harnessing discord could provide fundamental insights for resource-efficient quantum information processing, remarkably, even in the absence of entanglement.

In 1993, Bennet and coworkers presented a seminal quantum protocol of teleportation that transfers quantum information from a sender at one location to a receiver some distance away, while without physically moving the object itself[18]. At the heart of quantum teleportation lies the concept of quantum entanglement, as it serves as the so-called EPR channel for quantum state transfer distantly[19]. Quantum teleportation, serving as a key primitive across a variety of quantum information tasks, is able to overcome the distance limitation in quantum communication and to realize long-range interactions in quantum computation[20,21]. Here we resemble quantum teleportation to demonstrate a discord-enabled technique of optical imaging at a distance using a classical pseudo-thermal light source, in which discord, instead of entanglement, sustains the teleportation-like channel. Our experiment further reveals explicitly that the amount of discord, being determined by the degree of incoherence of pseudo-thermal light, is directly related to the quality of the formed image.

Our present scheme can be considered as a variant of our recently proposed theoretical protocol of quantum image teleportation using pseudo-thermal light[22]. Indeed, our experimental strategy is inspired by teleportation, but we would like to emphasize that here we just aim at presenting a new optical imaging method at a distance. More precisely, our scheme itself is rather than quantum state teleportation, as it does not work at the single-photon level; instead, it realizes how to mimic teleportation for quantum transmission of a classical image non-locally using pseudo-thermal light source.

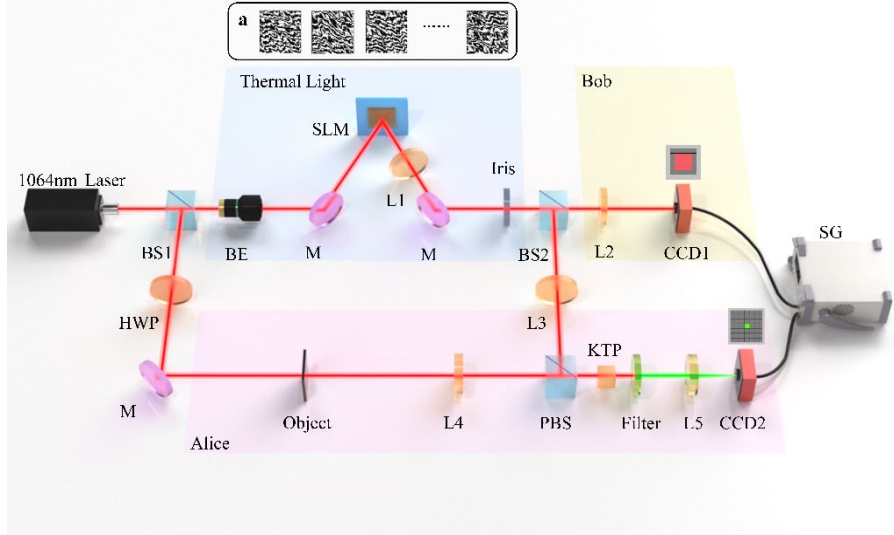

**Fig. 1 | Experimental setup for realizing the teleportation-like remote optical imaging with pseudo thermal light.** Inset **a** illustrates the frames of the dynamic hologram displayed on SLM for preparing the pseudo thermal light.

As sketched in Fig. 1, our optical setup follows the basic framework of teleportation protocol[18], in which discord offered by thermal light is used instead of entanglement between photon pairs. Conventionally, by guiding the coherent light beam to pass through dynamic Kolmogorov phase screens, we generate the pseudo-thermal light with tunable discord by modulating the transverse coherence width[23]. Then, the generated thermal light is divided to Alice and Bob with a beam splitter (BS). Based on the theory of randomly thermal radiation[24] and digital spiral imaging[25-27], thermal two-photon states shared by Alice and Bob can be expressed as[22],

$$\rho_{bc} = \rho_{bc}^0 + |\psi\rangle_{bc}\langle\psi|, \qquad (1)$$

with $\rho_{bc}^0 = \frac{1}{d}\sum_{\ell,p}|\ell,p\rangle_b\langle\ell,p|\sum_{\ell',p'}|\ell',p'\rangle_c\langle\ell',p'|$, which is essentially a diagonal state, representing a maximally mixed state. And $|\psi\rangle_{bc} = \sum_{\ell,p} P_{\ell,p}|\ell,p\rangle_b|-\ell,p\rangle_c$, which is inherently a pure entangled orbital angular momentum (OAM) state. Here, subscript *b* and *c* represent the photon *b* and *c* held by Alice and Bob, respectively, $\ell$ and *p*

represent the azimuthal and radial quantum number of Laguerre-Gaussian (LG) mode. Although the entire thermal two-photon state is separable, Eq. (1) reveals the presence of nonclassical correlations within it. This is also the reason why thermal light can be used to realize non-local correlated imaging, namely, ghost imaging[28-30]. In contrast to quantum correlated imaging, the first term in Eq. (1) contributes to a background noise. Also, being similar with the quantum correlated imaging[31,32], the imaging quality with thermal light can be controlled by adjusting the probability amplitude $P_{\ell,p}$. Particularly, it essentially captures the discrepancy between quantum mutual information and classical correlation in thermal bipartite system, namely, discord $D$ [33],

$$D(\rho) = \frac{\left(\sum_{\ell,p} P_{\ell,p}^2\right)^2 - \sum_{\ell,p} P_{\ell,p}^4}{\left(\sum_{\ell,p} P_{\ell,p}^2 + \left(\sum_{\ell,p} P_{\ell,p}\right)^2\right)^2}, \qquad (2)$$

and the probability amplitude $P_{\ell,p}$ is determined by the transverse size, $\sigma_s$, and coherence width, $\sigma_g$, of thermal light[22], that is,

$$P_{\ell,p} = \left(1 - \tan^4 \frac{\beta}{2}\right)\left(\tan^2 \frac{\beta}{2}\right)^{|\ell|+2p}, \qquad (3)$$

with $\tan\beta = 2\sigma_s/\sigma_g$. In quantum correlated imaging and teleportation, the prerequisite for perfect imaging and teleportation is the maximal OAM entanglement state, which is exactly the second term of Eq. (1) with an enough small coherence width. The discord of pure entangled state is

$$D(\rho_Q) = 1 - \frac{\sum_{\ell,p} P_{\ell,p}^4}{(\sum_{\ell,p} P_{\ell,p}^2)^2} \qquad (4)$$

And the imaging quality and teleportation fidelity degrade as the degree of entanglement decreases. Considering that the first term of Eq. (1) only contributes noise in the imaging process, quantum discord of the thermal two-photon source, embedded in the second term of Eq. (1), could provide a teleportation-like channel for realizing the optical imaging or optical image transport at a distance.

Assum Alice wishes to transport to Bob an optical image, $O(r,\varphi)$, which is carried by a coherent laser beam *a*. Drawing on the digital spiral imaging, the optical image can be expressed as, $|\psi\rangle_a = \sum_{\ell,p} A_{\ell,p} |\ell,p\rangle$, where the complex spiral spectrum $A_{\ell,p} = \int O(r,\varphi) \left[ \text{LG}_\ell^p (r,\varphi) \right]^* r dr d\varphi$, with $\text{LG}_\ell^p (r,\varphi)$ being the LG mode. As well known, in the framework of quantum teleportation, Alice needs to perform a Bell state measurement on photons *a* and *b*. However, performing the high-dimensional Bell state with linear optics remains a challenge. For this, we adopt the nonlinear sum-frequency generation process to realize the high-dimensional Bell-like state measurement[35]. Specifically, Alice mixes light beam *a* and thermal photons *b* in the Potassium titanyl phosphate (KTP) crystal to perform the nonlinear sum frequency generation (SFG). Considering the OAM conservation in SFG process and the Orthogonality of LG modes, if Alice merely detect the component of fundamental Gaussian mode, i.e., $\text{LG}_0^0 (r,\varphi)$, in SFG photons, Alice projects the light beam *a* and thermal photons *b* onto, $|\psi\rangle_{ab} = 1/\sqrt{d} \sum_{\ell,p} |\ell,p\rangle_a |-\ell,p\rangle_b$, namely, Alice realizes Bell-like state measurement. Subsequently, Alice informs Bob of her measurement results, namely, triggering the camera on Bob's side to capture the thermal photon *c*. Which can be expressed as,

$$\begin{aligned}\rho_c &= {}_{ab}\langle\psi|\rho_{abc}|\psi\rangle_{ab} \\ &= {}_{ab}\langle\psi|\left(|\psi\rangle_a\langle\psi|\rho_{bc}\right)|\psi\rangle_{ab} \\ &= \frac{1}{d}\sum_{\ell,p}|\ell,p\rangle_c\langle\ell,p| + |\phi\rangle_c\langle\phi|,\end{aligned} \quad (5)$$

with $|\phi\rangle_c = \sum_{\ell,p} A_{\ell,p} P_{\ell,p} |\ell, p\rangle$. From which we can see that the resultant quantum state of photon $c$ is a mixture of a completely mixed state and a pure state containing image information. Particularly, when the transverse coherence width is enough small, i.e., a completely incoherent thermal source, $P_{\ell,p}$ becomes a constant, corresponding to the discord value of 0. Thereby, Bob can acquire an exact replica of the original input image with a uniform noise background. In other words, by leveraging the thermal light source, we can mimic the teleportation to realize optical image transport between two separated parties without the direct transfer of information carriers. Meanwhile, the transport fidelity can be engineered by tailoring the discord of thermal two-photon source. Compared with the conventional ghost imaging with thermal light, our scheme represents another ghost imaging framework, therein, the photons involved in object illumination and image reconstruction come from entirely independent origins, namely, ghost imaging with interaction-free light.

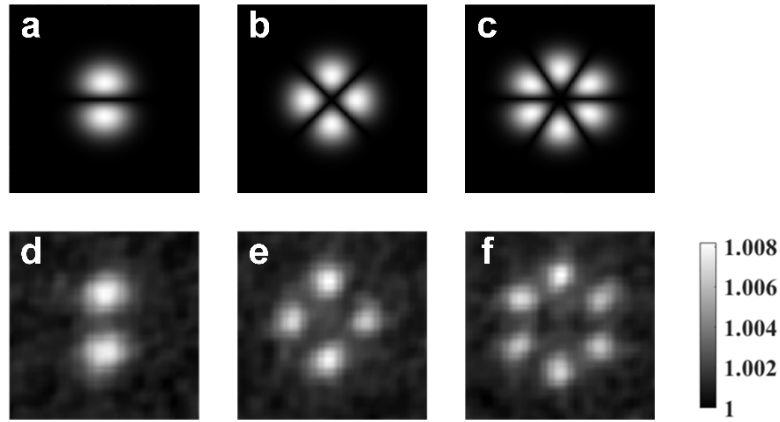

**Fig. 2 | Experimental observations for remote transport the OAM superposition states. a-c** The intensity distributions of the original superposition states $|\phi_1\rangle - |\phi_3\rangle$. **d-f** The corresponding transport results. Here, experimental results are acquired by summing 2000 frames.

To verify the effectiveness of the proposed teleportation-like optical image transport, we prepare three OAM superposition states to be transported at Alice side, $|\phi\rangle_1 = (|1\rangle + |-1\rangle)/\sqrt{2}$, $|\phi\rangle_2 = (|2\rangle + |-2\rangle)/\sqrt{2}$, and $|\phi\rangle_3 = (|3\rangle + |-3\rangle)/\sqrt{2}$. Due to the constructive and destructive interference along the angular direction, they exhibit a petal-shaped intensity distribution with the number of petals equal to $2|\ell|$, as indicated in the top row of Fig. 2. The implementation of the remote transporting is achieved through the coincidence measurement between the sum-frequency photon and the thermal photon *c*, see the Methods for details. The corresponding transported results are illustrated in the bottom row of Fig.2, from which we can clearly observe a petal-shaped intensity distribution, identical to that of the input state, emerging against a noisy background. This is in complete agreement with the theoretical predictions in Eq. (5), demonstrating that our scheme can be used for the remote transmission of orbital angular momentum superposition states.

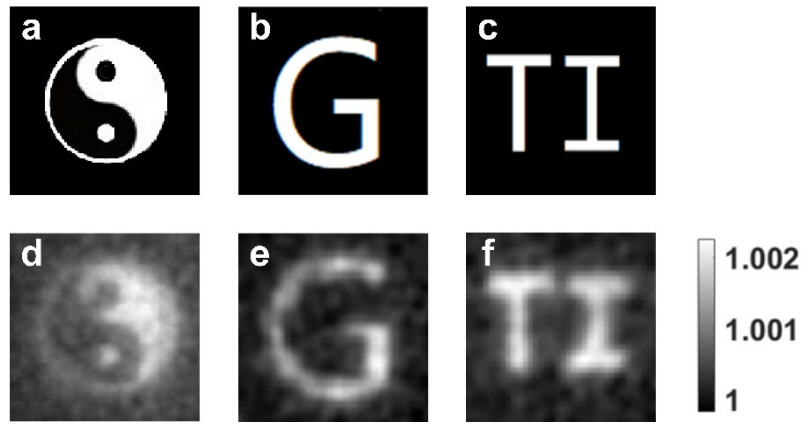

**Fig. 3 | Experimental observations for teleportation-like remote optical imaging. a-c** The original images of the Tai Chi, letters "G" and "TI", respectively. **d-f** The corresponding ghost images via remote transport. Here, the image of Tai Chi is acquired by summing 20000 frames, while the images of letters G and TI are acquired by summing 10000 frames.

Our teleportation-like remote transmission of two-dimensional OAM

superposition states can be extended to arbitrary dimensions, as predicted by Eq. (5). Furthermore, as aforementioned, based on the theory of digital spiral imaging, any two-dimensional image can be represented by OAM superposition states. Accordingly, our scheme can be generalized to achieve teleportation-like imaging with thermal light. Unlike traditional imaging, the photons used for illuminating the object and recovering the image do not interact with each other, namely, ghost imaging with interaction-free light[34,35]. This has not been explored in thermal light ghost imaging and will provide more possibilities for ghost imaging with thermal light. For this, as shown in Fig. 3a-c, we prepare the input images of Tai Chi, letters "G" and "TI" carried by the coherent light beam $a$ at Alice side. Then, the fundamental mode component of sum frequency photons is selected to serve as the trigger signal or correlated signal, which is used to trigger CCD 1 to capture thermal photons $c$. Here, the fundamental mode component is extracted by selecting the intensity signal from the central pixel of the sum-frequency light spot recorded by CCD2 in the Fourier plane. Notably, the pixel size of the CCD ($4\mu\text{m} \times 4\mu\text{m}$) is comparable to the core diameter of a single-mode fiber, ensuring precise mode selection. After the triggering or correlation operation, the transported images are acquired and shown in Fig. 3d–f. We employ the Contrast-to-Noise Ratio (CNR)[35,36], $\text{CNR} = (\langle G_{in} \rangle - \langle G_{out} \rangle)/\sqrt{\sigma_{in}^2 + \sigma_{out}^2}$, to quantitively assess the quality of acquired images, where $\langle G_{in} \rangle$ and $\langle G_{out} \rangle$ represent the ensemble average of the signal at any pixel inside and outside the image regions, respectively, and, $\sigma_{in}^2$ and $\sigma_{out}^2$ are the variances. After calculation, we obtain CNR= 2.7763, 2.6126 , 4.1567 for Fig. 3d, 3e, and 3f, respectively, which are comparable to the values in traditional ghost imaging with thermal light[36,37], thus confirming that our scheme enables efficient optical imaging between two distant parties without directly transmitting the image itself.

As indicated by the theoretical prediction by Eq. (5) and the experimental results in Fig. 2 and 3, we mimic teleportation to realize the remote transport of a classical image non-locally using pseudo-thermal light source, proposing a scheme for optical imaging at a distance. Although our scheme does not fall under the category of quantum teleportation, it shares similarity with quantum teleportation. That is remotely transporting classical image carried by the coherent beam to the distant without using knowledge of the state of the laser photons, which is a key feature of quantum teleportation. In quantum teleportation, entanglement provides the quantum channel for realizing this task. Theoretically, the fidelity of quantum teleportation deteriorates as the degree of entanglement decreases. Analogously, in our teleportation-like optical imaging with thermal light, discord provides a teleportation-like communication channel. As indicated by Eq. (5), the imaging quality also deteriorates as the quantum discord decreases. To verify this, following Eqs. (3) and (4), we sequentially vary the quantum discord of the thermal light source to $D(\rho_Q) = 0.8631, 0.9602, 0.9754, 0.9866, 0.994, 0.9964, 0.9981, 0.9992, 0.9997, 0.9999$, which is achieved by adjusting the dynamic turbulence phase loaded onto the SLM. For each discord value, we record the transported patterns of the letter TI and calculate their CNR. The corresponding results are illustrated in Fig. 4, which are consistent with the theoretical predications in Eq. (5). Therefore, it is demonstrated that the imaging performance of our teleportation-like optical imaging is dominated by discord.

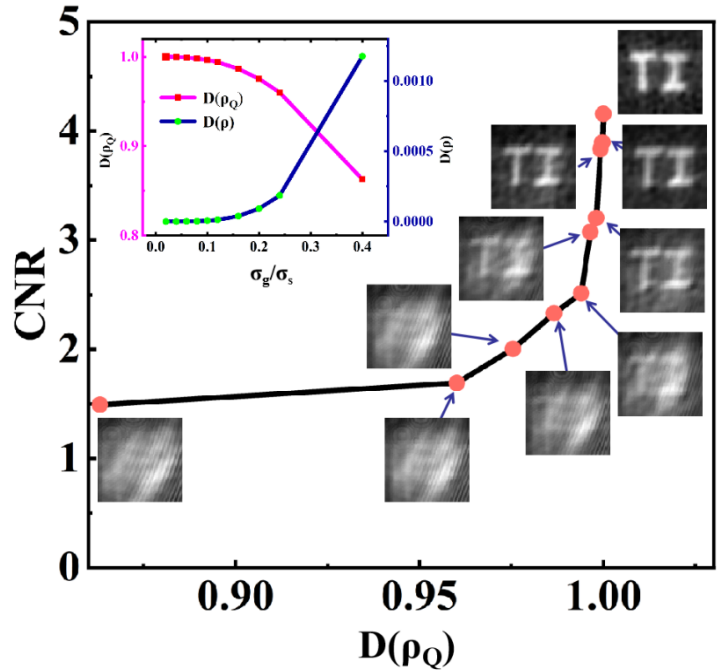

**Fig. 4 | Imaging performance of the teleportation-like remote optical imaging with different discord.** Insets represent the reconstructed images via remote transport with discord of $0.8631$, $0.9602$, $0.9754$, $0.9866$, $0.994$, $0.9964$, $0.9981$, $0.9992$, $0.9997$, $0.9999$, respectively. The black line represents the variation of CNR with discord. As a result of the different quantum states, the blue line describes discord of the thermal two-photon state, while the purple line corresponds to the discord of entangled state embedded in the thermal two-photon state.

**Discussion**

In summary, by leveraging the pseudo thermal light source, we constructed a teleportation-like channel. And by leveraging the SFG, we presented a scheme for high-dimensional Bell-like state measurement. Accordingly, we demonstrated the optical imaging between two distant parties without directly transporting image itself. We have experimentally achieved the remote transmission of OAM superposition states and two-dimensional intensity images. Also, analogous to the role of entanglement in quantum teleportation, we have elucidated the impact of discord of thermal light on imaging quality in our scheme. Compared to traditional correlated imaging, additional security guarantees are provided by encoding the information in a third light beam. Through Bell-like state measurements based on nonlinear optics, our proposal provides a potential to achieve cross-wavelength imaging, such as using infrared light for object illumination and visible light for imaging, thus overcoming the limitations of infrared detectors in ghost imaging. In our demonstration, we use a coherent laser to carry the information to be transported, it is anticipated to be extendable for information transfer between two or even multiple thermal light sources, enabling thermal light correlation swapping in a manner similar to entanglement swapping[34,38]. Furthermore, introducing the concept of quantum teleportation into thermal light ghost imaging will inspire further exploration into the fundamental nature of ghost imaging.

**Methods**

To achieve coincidence measurements between the sum-frequency photon and the thermal photon $c$, we employ an external level signal to trigger the two CCD cameras for synchronized data acquisition, rather than relying on traditional trigger photography. Specifically, a signal generator (SG) is used as the trigger source to emit square wave signals, and the CCDs are configured to respond to rising edge signals. When the voltage signal changed from 0 to 12 V, both CCDs capture images simultaneously and saved them to designated folders. Then, the coincidence measurements are realized through subsequent digital processing of the acquired images. Specifically, a second-order correlation operation between the speckle pattern $I_{1i}(r_1)$ captured by CCD1 at time $i$ and the central pixel value $I_{2i}(r_2=0)$ of the image captured by CCD2 can be calculated with,

$$G^{(2)}(r_1) = \frac{\frac{1}{n}\sum_{i}^{n} I_{1i}(r_1) I_{2i}(r_2=0)}{\frac{1}{n}\sum_{i}^{n} I_{1i}(r_1) \times \frac{1}{n}\sum_{i}^{n} I_{2i}(r_2=0)}, \qquad (6)$$

where $n$ represents the number of frames of the collected data, $r_1$ denotes the transvers coordinates in the plane of CCD1, which is placed at the image plane. Here, we select the intensity signal from the central pixel of the sum-frequency light spot recorded by CCD2 at the Fourier plane, i.e., $r_2=0$, to extracted the fundamental mode component of sum-frequency photon for ensuring the Bell-like measurement. And the transported image can be restored through the correlation operation of the two CCDs.

**Acknowledgments**

This work was supported by the National Natural Science Foundation of China (12034016, 61975169), the Natural Science Foundation of Fujian Province of China for Distinguished Young Scientists (2015J06002), and the program for New Century Excellent Talents in University of China (NCET-13-0495).

## Author contributions

L.C. conceived the idea and designed the experiment. Z.C., W.S. & X. Q. conducted the experiment with the help from Y. C. and L.C. All the authors analyzed the data and co-wrote the manuscript. L.C. supervised the project.

## Competing interests

The authors declare no competing interests.

## Additional information

**Supplementary information** The online version contains supplementary material available at

https://doi.org/10.1038/

**Correspondence and requests for materials** should be addressed to X. Q., Y. C. and L. C.

**Reprints and permissions information** is available at www.nature.com/reprints.